\documentclass[a4paper,11pt]{article}

\usepackage{jheppub} 

\title{\boldmath Thermal Interpretation of Schwinger Effect in Near-Extremal RN Black Hole}

\author[a,b,1]{Sang Pyo Kim}
\author[c]{Hyun Kyu Lee}
\author[c]{Yongsung Yoon}

\affiliation[a]{Department of Physics, Kunsan National University, Kunsan 573-701}
\affiliation[b]{Center for Relativistic Laser Science, Institute for Basic Science (IBS), Gwangju 500-712, Korea}
\affiliation[c]{Department of Physics, Research Institute for Natural Sciences, Hanyang University, Seoul 133-791, Korea}

\emailAdd{sangkim@kunsan.ac.kr}
\emailAdd{hyunkyu@hanyang.ac.kr}
\emailAdd{cem@hanyang.ac.kr}

\abstract{We propose a thermal interpretation of the Schwinger effect for charged scalars and spinors in an extremal and near-extremal Reissner-Nordstr\"{o}m (RN) black hole. The emission of charges has the distribution with an effective temperature determined by the Davies-Unruh temperature for accelerating charges by the electric field and the scalar curvature of ${\rm AdS}_2$ from the near-horizon geometry ${\rm AdS}_2 \times {\rm S}^2$. We find a charge bound for the extremal micro black hole to remain stable against the Schwinger emission in analogy with the Breitenlohlner-Freedman bound for the ${\rm AdS}$ space. In the in-out formalism we find the one-loop QED effective action consistent with the vacuum persistence and interpret the vacuum persistence as the leading Schwinger effect and the effect of a charged vacuum of the Coulomb field.}

\begin{document}
\maketitle
\flushbottom

\section{Introduction} \label{sec1}

A charged Reissner-Nordstr\"{o}m (RN) black hole has both an event horizon due to the mass and charge of the hole and a Coulomb field due to the charge.
The emission of particles from the RN black hole is thus characterized by the Hawking radiation, a thermal radiation of all species of particles with the Hawking temperature determined by the surface gravity on the horizon \cite{Hawking}, and by the Schwinger effect, production of charged pairs from the Dirac sea due to the electric field near the horizon \cite{Schwinger}. The Hawking radiation and the Schwinger effect are a nonperturbative quantum phenomenon, which cannot be obtained from a few Feynman diagrams. In a flat spacetime, a constant electric field spontaneously creates charged pairs from the Dirac sea according to the Boltzmann factor $e^{- \pi m^2/qE}$, which becomes efficient when the field strength is near the critical value of the rest mass of the pair.

The emission of particles from the RN black hole is an interesting issue because of the intertwinement of the Hawking radiation and the Schwinger effect, in other words, the electric field effect as well as the gravity effect. The Schwinger effect and the QED action may thus shed light on understanding the strong interaction of electromagnetic fields and gravity. However, the Schwinger effect has not been exactly calculated for the RN black hole except for the extremal or near-extremal limit. In particular, the Hawking radiation vanishes in an extremal black hole and is exponentially suppressed in a near-extremal black hole, so charges are emitted via the Schwinger mechanism due to the Coulomb field. The geometry ${\rm AdS}_2 \times {\rm S}^2$ near the horizon of the extremal or near-extremal black hole \cite{Bardeen99} may give an explicit formula for the Schwinger effect and the QED action since the QED action in a constant electric field has recently been found in the ${\rm dS}_2$ and ${\rm AdS}_2$ space \cite{Cai14}.

In this paper we advance a thermal interpretation of the emission of charged scalars and spinors from the extremal and near-extremal black hole and find the one-loop QED effective action. Recently, using the near-horizon geometry, Chen et al have found the emission rate of charged scalars \cite{Chen12} and spinors \cite{Chen14}. The Schwinger effect has been studied in charged black holes \cite{Zaumen,Carter,Damour,Gibbons,Page,Hiscock,Khriplovich,Gabriel,Kim05,Kim14} and also in charged black holes with an asymptotically ${\rm dS}$ or ${\rm AdS}$ space \cite{Belgiorno08,Belgiorno09,Belgiorno10}. Cai and one of the authors (S.P.K.) have proposed a thermal interpretation of the emission of charged scalars from a constant electric field in ${\rm dS}_2$ as well as ${\rm AdS}_2$ \cite{Cai14}. The pair production in ${\rm AdS}_2$ has the Breitenlohlner-Freedman (BF) bound due to the confining nature of virtual pairs by the ${\rm AdS}$ space \cite{Pioline05,Kim08a} in contrast to the enhancement in the ${\rm dS}$ space due to the Gibbons-Hawking radiation.

In the flat spacetime, the Schwinger pair production has a thermal interpretation in terms of the Davies-Unruh temperature \cite{Davies75,Unruh76} for an accelerating charge \cite{Brout91,Brout95,Hwang09,Labun12}. In the ${\rm dS}$ or ${\rm AdS}$ space, an accelerating observer has $T_{\rm eff} = \sqrt{T_{\rm U}^2 + {\cal R}/(8 \pi^2)}$, an effective temperature, of the Davies-Unruh temperature and the spacetime curvature \cite{Cai14,Narnhofer96,Deser97}, while the Schwinger formula has $T_{\rm eff} = T_{\rm U} + \sqrt{T_{\rm U}^2 + {\cal R}/(8 \pi^2)}$ with respect to an effective mass \cite{Cai14}. In the extremal black hole, the near-horizon geometry ${\rm AdS}_2 \times {\rm S}^2$ prescribes the effective temperature $T_{\rm AdS} = T_{\rm U} + \sqrt{T_{\rm U}^2 + {\cal R}_{\rm AdS}/(8 \pi^2)}$ with respect to another effective mass. In analogy with the BF bound in ${\rm AdS}$, the extremal black hole has a bound $T_{\rm U} < \sqrt{-{\cal R}_{\rm AdS}/(8 \pi^2)}$ and emits charged pairs via the Schwinger mechanism when this bound is violated. This implies that extremal micro black holes can be stable against the Hawking radiation and the Schwinger effect.

In the in-out formalism by Schwinger and DeWitt \cite{DeWitt75,DeWitt03}, the one-loop effective action is expressed as the logarithm of the Bogoliubov coefficient modulo renormalization of the vacuum energy and charge. The one-loop QED effective action has been found for pulsed or localized electric fields in the in-out formalism \cite{Kim08b,Kim10,Kim12a,Kim12b}.  From the Bogoliubov transformation, we find the one-loop QED action on the horizon of the extremal or near-extremal black hole. The effective action is complex and the vacuum persistence, twice the imaginary of the action, is determined by the mean number of produced pairs. In fact, the one-loop QED action is expressed as a sum of Schwinger's proper-time integrals. The QED action for scalars and spinors has the form of the standard QED action with an additional contribution from the charged vacuum of Coulomb field.

The organization of this paper is as follows. In section \ref{sec2} we advance a thermal interpretation of the leading Schwinger effect for an extremal and near-extremal RN black hole. The leading term of the Schwinger formula is derived from the Hamilton-Jacobi action via the phase-integral method. We find the bound for black hole charge against the charge emission. In section \ref{sec3} we propose the thermal interpretation of the exact Schwinger formula for the extremal and near-extremal black hole and compare it with that of a constant field in ${\rm AdS}_2$. We study the vacuum persistence due to the charge emission and charged vacuum. Using the gamma-function regularization, we find the one-loop QED effective action as a sum of Schwinger's proper-time integrals. In section \ref{sec4} we discuss the physical implication of the Schwinger effect in charged black holes.

\section{Effective Temperature for Leading Schwinger Effect in RN Black Hole}\label{sec2}

The motion of a charged scalar with the mass $m$ and charge $q$ in the four-vector potential $A_{\mu}$ for an electromagnetic field and in a curved spacetime $g_{\mu \nu}$ obeys the Klein-Gordon equation (in the Planck units of $c = \hbar = G = k_B = 1/(4 \pi \epsilon_0)= 1$)
\begin{eqnarray}
\frac{1}{\sqrt{-g}} \hat{\pi}_{\mu} \Bigl(\sqrt{-g} g^{\mu \nu}  \hat{\pi}_{\nu} \Bigr) \phi - m^2 \phi = 0, \label{KG eq}
\end{eqnarray}
where $\hat{\pi}_{\mu} = \hat{p}_{\mu} - q A_{\mu}$. The charged RN black hole with the metric
\begin{eqnarray}
ds^2 = - \Bigl(1 - \frac{2M}{r} + \frac{Q^2}{r^2} \Bigr) dt^2 + \frac{dr^2}{1 - \frac{2M}{r} + \frac{Q^2}{r^2}} + r^2 d \Omega_2^2,
\end{eqnarray}
under a mapping $r - Q= \epsilon \rho, t = \tau/\epsilon$ and a reparametrization $M-Q = (\epsilon B)^2/(2Q)$, has the near-horizon geometry of ${\rm AdS}_2 \times {\rm S}^2$ \cite{Chen12}
\begin{eqnarray}
ds^2 = - \frac{\rho^2 - B^2}{Q^2} d\tau^2 + \frac{Q^2}{\rho^2 - B^2} d \rho^2 + Q^2 d \Omega_2^2. \label{AdS}
\end{eqnarray}
Then, the Hawking temperature on the horizon at $r_{H} = Q + \epsilon B$ $(\rho =B)$ is given by
\begin{eqnarray}
T_{\rm H} = \frac{\sqrt{\frac{1}{2} + \frac{M}{2Q}}}{4 \pi \Bigl( M+ \sqrt{\frac{1}{2} + \frac{M}{2Q}} \epsilon B \Bigr)^2} \epsilon B, \label{Hawking}
\end{eqnarray}
and the Coulomb potential of the black hole is $A_0 = - \rho/Q$ and the electric field on the horizon is $E_{H} =1/Q$.

The leading term for the emission of the same kind of charges from the near-extremal RN black hole can be determined by the Hamilton-Jacobi action of the spherically symmetric form ($qQ>0$)
\begin{eqnarray}
\phi = e^{i S (\rho) - i \omega \tau} Y_{lm} (\theta, \varphi), \label{wave}
\end{eqnarray}
in which the action from eq. (\ref{KG eq}) is given by
\begin{eqnarray}
S (\rho) = \int \frac{d \rho}{\rho^2 - B^2} \sqrt{(q \rho - \omega Q)^2 Q^2 - \Bigl(m^2 Q^2 + (l+ \frac{1}{2})^2 \Bigr) (\rho^2 - B^2)}. \label{HJ action}
\end{eqnarray}
The particle state is well defined by the wave function (\ref{wave}) for $\rho \gg 1$ in the near horizon provided that the bound\footnote{All the masses, lengths and charges are measured in the Planck units, $m_p$, $l_p$ and $q_p$. The Planck charge $q_p = \sqrt{4 \pi \epsilon_0 \hbar c}$ is given by $q_p = \frac{e}{\sqrt{\alpha}}$ from the fine structure constant $\frac{1}{4 \pi \epsilon_0}  \frac{e^2}{\hbar c} = \alpha$. We may write eq. ({\ref{BF bound}) as $Q \leq \bigl(J+ \frac{1}{2} \bigr) q_m \bigl( 1 - \frac{m^2}{q^2} \bigr)^{-1/2}$} with the Dirac monopole charge $q_m = \frac{e}{\alpha}$ and the angular momentum $J$ ($J = 0, 1, \cdots$ for scalars and $J = 1/2, 3/2, \cdots$ for spinors).}
\begin{eqnarray}
Q \leq \sqrt{\frac{(l + \frac{1}{2})^2}{q^2 - m^2}} \label{BF bound}
\end{eqnarray}
is violated, so the black hole satisfying the bound (\ref{BF bound}) cannot produce pairs of charged scalars. A few comments are in order. First, the bound (\ref{BF bound}) is the black hole analog of the BF bound in ${\rm AdS}$ \cite{Cai14,Pioline05,Kim08a}. Second, the bound is entirely determined by the charge, mass and angular momentum of the emitted particle. Third, both the Hawking radiation and the Schwinger mechanism are prohibited for the extremal black hole with the charge (\ref{BF bound}) for that particular scalars and the extremal micro black hole satisfying the charge bound for all species of charged particles can remain a stable object.

Using the tunneling boundary condition for QED, the leading contribution to Schwinger formula can be obtained by the phase-integral formula \cite{Kim07}
\begin{eqnarray}
N_{\rm S} = e^{i {\cal S}_{\Gamma}},
\end{eqnarray}
where ${\cal S}_{\Gamma}$ is the action evaluated along a contour $\Gamma$ in the complex plane of $\rho$. A similar formula  in the complex plane of time holds for pair production from a time-dependent electric field or an expanding spacetime \cite{Kim13a,Kim13b}. In the complex plane of $\rho$, the contour integral (\ref{HJ action}) has a pair of simple poles at $\rho = \pm B$ and another simple pole at $\rho = \infty$, whose residues contribute ${\cal S}_{a}$ and ${\cal S}_{b}$, respectively, to the Schwinger formula
\begin{eqnarray}
N_{\rm S} = e^{- {\cal S}_{a} + {\cal S}_{b} }, \label{pair prod}
\end{eqnarray}
where
\begin{eqnarray}
{\cal S}_{a} = 2 \pi q Q, \quad {\cal S}_{b} = 2 \pi \sqrt{(q^2 - m^2) Q^2 - (l+ \frac{1}{2})^2 }. \label{instanton}
\end{eqnarray}
From now on, we shall consider a positive instanton action ${\cal S}_{b}$, which violates the bound (\ref{BF bound}) and thus emits charged scalars.

The Davies-Unruh temperature for an accelerating charge near the horizon is
\begin{eqnarray}
T_{\rm U} = \frac{qE_H/\bar{m}}{2 \pi} = \frac{q}{2 \pi \bar{m} Q},
\end{eqnarray}
where $\bar{m}$ is an effective mass,
\begin{eqnarray}
\bar{m} = m \sqrt{1+ \Bigl(\frac{l+ \frac{1}{2}}{mQ}\Bigr)^2}. \label{mass}
\end{eqnarray}
Here, the angular momentum effectively adds to the mass of charged scalars in eq. (\ref{KG eq}).
Then, the Schwinger formula (\ref{pair prod}) takes the form
\begin{eqnarray}
N_{\rm S} = e^{- \frac{\bar{m}}{T_{\rm RN}} }, \label{Boltzmann}
\end{eqnarray}
where $T_{\rm RN}$ is the effective temperature on the horizon:
\begin{eqnarray}
T_{\rm RN} = T_{\rm U} + \sqrt{T_{\rm U}^2 - \Bigl( \frac{1}{2 \pi Q} \Bigr)^2}. \label{eff tem}
\end{eqnarray}
Noting that ${\cal R}_{\rm AdS} = -2/Q^2$ for the metric (\ref{AdS}), the effective mass (\ref{mass}) for an $s$-wave and the effective temperature (\ref{eff tem}) take the same form as QED in a constant electric field in ${\rm AdS}_2$ \cite{Cai14}:
\begin{eqnarray}
\bar{m}_{\rm AdS} = m \sqrt{ 1 - \frac{{\cal R}_{\rm AdS}}{8 m^2}}, \quad T_{\rm AdS} = T_{\rm U} + \sqrt{T_{\rm U}^2  + \frac{{\cal R}_{\rm AdS}}{8 \pi^2}}
\end{eqnarray}
where $T_{\rm U} = (qE/\bar{m}_{\rm AdS})/(2 \pi)$ is the Davies-Unruh temperature.

\section{Schwinger Effect, Vacuum Persistence and One-Loop QED Action  in RN Black Hole} \label{sec3}

In the in-out formalism by Schwinger and DeWitt, the one-loop effective action is obtained from the Bogoliubov coefficient \cite{DeWitt75,DeWitt03}
\begin{eqnarray}
W = \int d^4x {\cal L}_{eff} = \pm i \sum_{J} (2J+1) \ln(\alpha^* (J)), \label{SD action}
\end{eqnarray}
where the upper (lower) sign is for scalars (spinors) and the angular quantum number $J$ runs $J=0, 1, \cdots$ for scalars and $J = 1/2, 3/2, \cdots$ for spinors. Thus, the vacuum persistence, the twice of the imaginary part of the effective action,
\begin{eqnarray}
2 {\rm Im} W = \pm \sum_{J} (2J+1) \ln(1 \pm N (J)), \label{vac per}
\end{eqnarray}
is related to the mean number of produced pairs and prescribes the probability for the vacuum to remain in the in-vacuum, in other words, the decay probability of the vacuum due to the emission of pairs. The vacuum persistence may be interpreted as the pressure of ideal quantum gas of scalars or spinors \cite{Ritus}. The repulsive quantum nature of spinors enhances the probability for the vacuum decay. To calculate the mean number, the vacuum persistence and the one-loop effective action, we use the Bogoliubov coefficients for the Klein-Gordon equation and the Dirac equation in the geometry (\ref{AdS}) in the extremal and near-extremal RN black holes \cite{Chen12,Chen14}.

\subsection{Schwinger Effect in Extremal RN Black Hole}\label{sec3-1}

The extremal RN black hole has the zero surface gravity and the Hawking radiation is thus exponentially suppressed to zero. The vacuum persistence amplitude for scalars and spinors is given by refs. \cite{Chen12,Chen14}
\begin{eqnarray}
|\alpha_{\rm EBH}|^2 = \frac{1 \pm e^{- {\cal S}_{a} + {\cal S}_{b}}}{1 \pm e^{- {\cal S}_{a} - {\cal S}_{b} }}, \label{vac per}
\end{eqnarray}
where the upper (lower) sign is for scalars (spinors) and
\begin{eqnarray}
{\cal S}_{b} = 2 \pi qQ \sqrt{1 - \Bigl(\frac{m}{q} \Bigr)^2 \Bigl(1 + \frac{(J+ \frac{1}{2})^2}{(mQ)^2} \Bigr)},
\end{eqnarray}
where $J = 0, 1, \cdots$ for scalars and $J = 1/2, 3/2, \cdots$ for spinors.
The mean number of emitted scalars and spinors, simply the Schwinger formula, follows from refs. \cite{Chen12,Chen14}
\begin{eqnarray}
N_{\rm EBH} = \pm (|\alpha_{\rm EBH}|^2 -1) = \frac{e^{- {\cal S}_{a} + {\cal S}_{b}} - e^{- {\cal S}_{a} - {\cal S}_{b}}}{1 \pm e^{- {\cal S}_{a} - {\cal S}_{b} }}. \label{Schwinger}
\end{eqnarray}

We now propose a thermal interpretation for the Schwinger formula (\ref{Schwinger}). Introducing another parameter associated with  $R_{\rm RN}$
\begin{eqnarray}
{\cal T}_{\rm RN} = T_{\rm U} - \sqrt{T_{\rm U}^2 - \Bigl( \frac{1}{2 \pi Q} \Bigr)^2}, \label{eff tem-2}
\end{eqnarray}
the Schwinger formula (\ref{Schwinger}) can be expressed as
\begin{eqnarray}
N_{\rm EBH} = \frac{e^{- \frac{\bar{m}}{T_{\rm RN}}} - e^{- \frac{\bar{m}}{{\cal T}_{\rm RN}}} }{1 \pm e^{- \frac{\bar{m}}{{\cal T}_{\rm RN}}}}, \label{therm pair}
\end{eqnarray}
where the effective mass
\begin{eqnarray}
\bar{m} = m \sqrt{1+ \Bigl(\frac{J+ \frac{1}{2}}{mQ}\Bigr)^2}. \label{mass-2}
\end{eqnarray}
Note that
\begin{eqnarray}
T_{\rm RN} {\cal T}_{\rm RN} = \Bigl( \frac{1}{2 \pi Q} \Bigr)^2 = - \frac{{\cal R}_{\rm AdS}}{8 \pi^2}. \label{tem dual}
\end{eqnarray}
The emission is dominated by the effective temperature $T_{\rm RN} $ because  $T_{\rm RN} > {\cal T}_{\rm RN}$ and $e^{- \bar{m}/{\cal T}_{\rm RN}}$ gives correction terms to the Boltzmann factor (\ref{Boltzmann}). Note that the vacuum persistence amplitude (\ref{vac per}) and
the Schwinger formula (\ref{therm pair}) for scalars are exactly the same as eqs. (4.10) and (4.11) in ${\rm AdS}_2$ \cite{Cai14}. The formula with the lower sign is new result for spinors in ${\rm AdS}_2$.

The thermal interpretation for the Schwinger formula (\ref{therm pair}) may be understood from the vacuum persistence
\begin{eqnarray}
2 {\rm Im} W = \pm \Bigl( \ln (1 \pm e^{-\frac{\bar{m}}{T_{\rm RN}}}) - \ln (1 \pm e^{-\frac{\bar{m}}{\tilde{\cal T}_{\rm RN}}}) \Bigr). \label{vac per-ex}
\end{eqnarray}
The first logarithm in eq. (\ref{vac per-ex}) is the standard QED action with the mean number $N{\rm S} = e^{-\bar{m}/T_{\rm RN}}$ for scalars and spinors \cite{Schwinger} while the second one is a correction due to a charged vacuum in the Coulomb field. In a flat spacetime, a constant electric field uniformly tilts the Dirac vacuum, from which a particle from near the negative Fermi energy tunnels quantum mechanically to create a pair with $N_{\rm S} = e^{- \pi m^2/qE}$ of particle and antiparticle as a hole. On the other hand, in a supercritical Coulomb field the energy of binding states can be lower than the Dirac vacuum, and particles with the opposite charge may be confined to the bound states while antiparticles with the same charge escape to infinity due to the Coulomb repulsion and form a charged vacuum \cite{Rafelski74}. In fact, the inhomogeneity of the Coulomb field changes the Dirac vacuum, which in turn modifies the vacuum persistence and subtracts the second term from the first one \cite{Kim14}. The vacuum persistence in a localized electric field exhibits such a feature in refs. \cite{Kim08b,Kim10}.

\subsection{Schwinger Effect in Near-Extremal RN Black Hole}\label{sec3-2}

The near-extremal RN black hole has small Hawking temperature (\ref{Hawking}). The vacuum persistence amplitude from refs. \cite{Chen12,Chen14} is given by
\begin{eqnarray}
|\alpha_{\rm NBH}|^2 = \Biggl( \frac{1 \pm e^{- {\cal S}_{a} + {\cal S}_{b}}}{1 \pm e^{- {\cal S}_{a} - {\cal S}_{b} }} \Biggr) \Biggl(\frac{1+e^{- \tilde{\cal S}_{a} - {\cal S}_{b}}}{1+ e^{- \tilde{\cal S}_{a} + {\cal S}_{b} }} \Biggr), \label{vac per-n}
\end{eqnarray}
where the upper (lower) sign is for scalars (spinors) and $\tilde{\cal S}_{a}$ is another instanton action
\begin{eqnarray}
\tilde{\cal S}_{a}= 2 \pi \frac{\omega Q^2}{B}.
\end{eqnarray}
Note that the second parenthesis in eq. (\ref{vac per-n}) is the additional common factor to eq. (\ref{vac per}).
The Schwinger formula also from refs. \cite{Chen12,Chen14} is
\begin{eqnarray}
N_{\rm NBH} =  \Biggl( \frac{e^{- {\cal S}_{a} + {\cal S}_{b}} - e^{- {\cal S}_{a} - {\cal S}_{b}}}{1 \pm e^{- {\cal S}_{a} - {\cal S}_{b} }}\Biggr) \Biggl(\frac{1 \mp e^{- \tilde{\cal S}_{a} + {\cal S}_{a}}}{1+ e^{- \tilde{\cal S}_{a} + {\cal S}_{b} }} \Biggr). \label{Schwinger-n}
\end{eqnarray}
The first parenthesis is the Schwinger formula (\ref{Schwinger}) in the extremal black hole while the second parenthesis is entirely due to the near extremity of black hole.

To have a thermal interpretation of the Schwinger formula (\ref{Schwinger-n}), we may introduce another pair of parameters associated to $T_{\rm RN}$ as
\begin{eqnarray}
\frac{2}{T_{\rm M}} =  \frac{1}{{\cal T}_{\rm RN}}+ \frac{1}{T_{\rm RN}}, \quad \frac{2}{{\cal T}_{\rm M}} = \frac{1}{{\cal T}_{\rm RN}} - \frac{1}{T_{\rm RN}}.
\end{eqnarray}
Then, the Schwinger formula takes the form
\begin{eqnarray}
N_{\rm NBH} = \Biggl( \frac{e^{- \frac{\bar{m}}{T_{\rm RN}}} - e^{- \frac{\bar{m}}{{\cal T}_{\rm RN}}} }{1 \pm e^{- \frac{\bar{m}}{{\cal T}_{\rm RN}}}} \Biggr)
\Biggl(\frac{1 \mp e^{- (\frac{\omega_t}{T_{\rm H}} - \frac{\bar{m}}{T_{\rm M}})}}{1 + e^{- (\frac{\omega_t}{T_{\rm H}} - \frac{\bar{m}}{{\cal T}_{\rm M}})}} \Biggr), \label{therm pair-n}
\end{eqnarray}
where $\omega_t = \omega \epsilon$ is the energy measured in the time $t$ in the asymptotically flat region. The emission is dominated by the Schwinger effect with the temperature $T_{\rm RN}$ while the Hawking radiation is a small correction with the Hawking temperature (\ref{Hawking}). A passing remark is that the emission (\ref{therm pair-n}) of spinors and the vacuum persistence (\ref{vac per-n}) is more efficient than those for scalars due to the quantum repulsion of spinor gas.

\subsection{One-Loop QED Action and Vacuum Persistence}\label{sec3-3}

The one-loop effective action (\ref{SD action}) is complex due to the emission of charged particles.
In the near-extremal black hole the Bogoliubov coefficient for scalars from ref. \cite{Chen12} is
\begin{eqnarray}
\alpha_{sc} (l) =  \Biggl[ \Biggl(\frac{\Gamma(\frac{1}{2} + i \frac{{\cal S}_a}{2 \pi} + i \frac{{\cal S}_b}{2 \pi} )}{\Gamma(\frac{1}{2} + i \frac{{\cal S}_a}{2 \pi} - i \frac{{\cal S}_b}{2 \pi} )} \Biggr) \Biggl(\frac{\Gamma(\frac{1}{2} - i \frac{\tilde{\cal S}_a}{2 \pi} + i \frac{{\cal S}_b}{2 \pi} )}{\Gamma(\frac{1}{2} - i \frac{\tilde{\cal S}_a}{2 \pi} - i \frac{{\cal S}_b}{2 \pi} )} \Biggr) \Biggr] \Biggl[\frac{\Gamma(-i 2 {\cal S}_b)}{\Gamma(i2 {\cal S}_b)} \Biggr], \label{a sc}
\end{eqnarray}
and for spinors from ref. \cite{Chen14} is
\begin{eqnarray}
\alpha_{sp} (j) = \Biggl[ \Biggl(\frac{\Gamma(1 + i \frac{{\cal S}_a}{2 \pi} + i \frac{{\cal S}_b}{2 \pi} )}{\Gamma(1 + i \frac{{\cal S}_a}{2 \pi} - i \frac{{\cal S}_b}{2 \pi} )} \Biggr) \Biggl(\frac{\Gamma(\frac{1}{2} - i \frac{\tilde{\cal S}_a}{2 \pi} + i \frac{{\cal S}_b}{2 \pi} )}{\Gamma(\frac{1}{2} - i \frac{\tilde{\cal S}_a}{2 \pi} - i \frac{{\cal S}_b}{2 \pi} )} \Biggr) \Biggr] \Biggl[\sqrt{\frac{{\cal S}_a-{\cal S}_b}{{\cal S}_a+{\cal S}_b}}\frac{\Gamma(1 -2i {\cal S}_b)}{\Gamma(1+ 2 i {\cal S}_b)} \Biggr]. \label{a sp}
\end{eqnarray}
The Bogoliubov coefficient in the extremal black hole is the limiting case of $\tilde{\cal S}_{a} = \infty$, for which the second parenthesis becomes unity. Thus, we restrict the vacuum persistence and the one-loop action in the near-extremal black hole.

Using the gamma-function regularization for the effective action (\ref{SD action}) in refs. \cite{Cai14,Kim08b,Kim10,Kim12a,Kim12b}, we perform one contour integral of the gamma functions in the first parenthesis along an infinite quarter circle in the first quadrant and another contour integral of the gamma functions in the second parenthesis along an infinite quarter circle in the fourth quadrant. We then obtain the real part of the renormalized one-loop effective action (vacuum polarization) for scalars
\begin{eqnarray}
W_{sc} = - \sum_{l} (2l+1) {\cal P} \int_{0}^{\infty} \frac{ds}{s} e^{-\frac{{\cal S}_{a} s}{2\pi}} \sinh \Bigl(\frac{{\cal S}_{b}s }{2\pi} \Bigr) \Biggl[ \frac{1}{\sin(\frac{s}{2})} - \frac{2}{s} - \frac{s}{12} \Biggr]\nonumber\\
- \sum_{l} (2l+1) {\cal P} \int_{0}^{\infty} \frac{ds}{s} e^{-\frac{\tilde{\cal S}_{a} s}{2\pi}} \sinh \Bigl(\frac{{\cal S}_{b} s}{2\pi} \Bigr) \Biggl[\frac{1}{\sin(\frac{s}{2})} - \frac{2}{s} - \frac{s}{12} \Biggr], \label{pol sc}
\end{eqnarray}
and for spinors
\begin{eqnarray}
W_{sp} = \sum_{j} (2j+1) {\cal P} \int_{0}^{\infty} \frac{ds}{s} e^{-\frac{{\cal S}_{a} s}{2\pi}} \sinh \Bigl(\frac{{\cal S}_{b} s}{2\pi} \Bigr) \Biggl[\frac{\cos(\frac{s}{2})}{\sin(\frac{s}{2})} - \frac{2}{s} + \frac{s}{6} \Biggr]\nonumber\\
+ \sum_{j} (2j+1) {\cal P} \int_{0}^{\infty} \frac{ds}{s} e^{-\frac{\tilde{\cal S}_{a}s}{2\pi}} \sinh \Bigl(\frac{{\cal S}_{b}s}{2\pi} \Bigr) \Biggl[\frac{1}{\sin(\frac{s}{2})} - \frac{2}{s} - \frac{s}{12} \Biggr]. \label{pol sp}
\end{eqnarray}
Here ${\cal P}$ denotes the principal value prescription and the terms subtracted from the ${\rm cosec}(s/2)$ and $\cot(s/2)$ corresponds to renormalization of the vacuum energy and charge. The imaginary part of the effective action is the sum of all the residues of simple poles along the imaginary axis in the contour integrals, and the vacuum persistence is thus given by
\begin{eqnarray}
2 {\rm Im} W &=&  \pm \sum_{J} (2J+1) \ln \Biggl[  \frac{(1 \pm e^{- {\cal S}_{a} + {\cal S}_{b}})(1 + e^{- \tilde{\cal S}_{a} - {\cal S}_{b}}) )}{(1 \pm e^{- {\cal S}_{a} - {\cal S}_{b}})(1 + e^{- \tilde{\cal S}_{a} + {\cal S}_{b}})} \Biggr] \nonumber\\
&=& \pm \sum_{J} (2J+1) \ln (|\alpha_{\rm NBH} (J)|^2). \label{W}
\end{eqnarray}
The gravitation attraction reduces (\ref{W}) for scalars in the near-extremal black holes while the quantum repulsion for spinors enhances (\ref{W}).

The vacuum polarization and persistence for the extremal black hole with $B=0$ in eq. (\ref{AdS}) is the limiting case of $\tilde{\cal S}_{a}= \infty$. The scalar vacuum polarization (\ref{pol sc}) for the $s$-wave is exactly the same as eq. (4.8) of ref. \cite{Cai14} while the spinor vacuum polarization in ${\rm AdS}_2$ is a new result:
\begin{eqnarray}
{\cal L}_{sp} = \frac{qE}{2 \pi} {\cal P} \int_{0}^{\infty} \frac{ds}{s} e^{-\frac{{\cal S}_{\kappa} s}{2\pi}} \sinh \Bigl(\frac{{\cal S}_{\nu} s}{2\pi} \Bigr) \Biggl[\frac{\cos(\frac{s}{2})}{\sin(\frac{s}{2})} - \frac{2}{s} + \frac{s}{6} \Biggr], \label{ads pol sp}
\end{eqnarray}
where
\begin{eqnarray}
{\cal S}_{\kappa} = -\frac{4 \pi qE}{{\cal R}_{\rm AdS}}, \quad {\cal S}_{\nu} = - \frac{4 \pi}{{\cal R}_{\rm AdS}} \sqrt{(qE)^2 + \frac{{\cal R}_{\rm AdS}}{2} \Bigl(m^2 - \frac{{\cal R}_{\rm AdS}}{2} \Bigr) }.
\end{eqnarray}
The vacuum polarization (\ref{ads pol sp}) is the spinor QED action except for the additional factor $\sinh ({\cal S}_{\nu} s/(2\pi) )$ due to the charged vacuum of the Coulomb field as explained in section \ref{sec3-1}.

\section{Conclusion}\label{sec4}

We have proposed a thermal interpretation of the emission of charged scalars and spinors from the extremal and near-extremal RN black holes. The leading term (\ref{pair prod}) of the Schwinger formula in the extremal black hole has the Boltzmann factor (\ref{Boltzmann}) with an effective mass and with an effective temperature (\ref{eff tem}) determined by the Davies-Unruh temperature and the scalar curvature of ${\rm AdS}_2$ with the spherical part adding angular momentum to the effective mass. We have shown that because of the near-horizon geometry of ${\rm AdS}_2 \times {\rm S}^2$ the one-to-one correspondence of the Schwinger formula exists between the extremal black hole and the ${\rm AdS}_2$ space. The extremal black hole has a bound of charge (\ref{BF bound}) for stability against the Schwinger emission, which is analogous to the Breitenlohlner-Freedman (BF) bound for a constant electric field in the ${\rm adS}$ space. We argue that extremal micro black holes within the charge bound for any species of charged particles may remain a stable object against both the Hawking radiation and the Schwinger emission. The thermal interpretation of the Schwinger formula for the near-extremal black hole is also advanced.

In the in-out formalism we have found the one-loop QED effective action (vacuum polarization) for scalars and spinors in the extremal and near-extremal black holes. We have shown that the scalar vacuum polarization for the $s$-wave in the extremal black hole is exactly the QED action in a constant electric field in ${\rm AdS}_2$, and have derived the spinor QED action in ${\rm AdS}_2$. The vacuum persistence, twice of the imaginary part of the effective action, is determined by the mean number of pairs produced in consistency with the in-out formalism. We have also shown that the vacuum persistence has the standard form for QED action for scalars and spinors and that the charged vacuum in the Coulomb field of the black hole reduces the vacuum persistence. The emission of charged particles in the near-extremal black hole is factorized into the emission in the extremal black hole and an additional factor that is dominated by the Hawking radiation. The Schwinger mechanism (mean number and vacuum persistence) for spinors is more efficient than that for scalars due to the repulsive quantum nature of spinors in contrast to the Hawking radiation.

The thermal interpretation of the emission from the near-extremal black hole may have an implication to the thermalization of quark-gluon plasma with the Davies-Unruh temperature \cite{Castorina07}. The Hawking radiation and the Schwinger emission from non-extremal black holes, which goes beyond the scope of this paper, will be important in understanding the evolution of charged black holes through the evaporation and discharge \cite{Hiscock90}. The holographic Schwinger effect and the AdS/CFT correspondence will be addressed in future publication.

\acknowledgments
The authors would like to thank Rong-Gen Cai and Chiang-Mei Chen for useful discussions. This work was supported in part by Basic Science Research Program through the National Research Foundation of Korea (NRF) funded by the Ministry of Education (NRF-2012R1A1B3002852) and in part by IBS (Institute for Basic Science) under IBS-R012-D1.

\end{document}